\documentclass[5p,times]{elsarticle}
\PassOptionsToPackage{hyphens}{url}\usepackage{hyperref}
\usepackage{multirow}
\usepackage{booktabs}
\usepackage{amsmath,amsfonts}
\usepackage{algorithmic}
\usepackage{algorithm}
\usepackage{array}
\usepackage[table, dvipsnames]{xcolor}
\usepackage[caption=false,font=normalsize,labelfont=sf,textfont=sf]{subfig}
\usepackage{textcomp}
\usepackage{stfloats}
\usepackage{verbatim}
\usepackage{graphicx}
\usepackage{breqn}  
 
\usepackage{rotating}

\begin{document}
\begin{sloppypar}

\begin{frontmatter}
\title{DI-NIDS: Domain Invariant \\Network Intrusion Detection System}

\cortext[cor]{Corresponding author}
\author[1]{Siamak Layeghy\corref{cor}}
\ead{siamak.layeghy@uq.net.au}

\author[1]{Mahsa Baktashmotlagh}
\ead{m.baktashmotlagh@uq.edu.au}

\author[1]{Marius Portmann}
\ead{marius@ieee.org}

\address[1]{School of ITEE, The University of Queensland, Brisbane, Australia}

\begin{abstract}
The performance of machine learning based network intrusion detection systems (NIDSs) severely degrades when deployed on a network with significantly different feature distributions from the ones of the training dataset.
In various applications, such as computer vision, domain adaptation techniques have been successful in mitigating the gap between the distributions of the training and test data. In the case of network intrusion detection however, the state-of-the-art domain adaptation approaches have had limited success. According to recent studies, as well as our own results, the performance of an NIDS considerably deteriorates when the `unseen' test dataset does not follow the training dataset distribution.
%
%
In order to enhance the generalisibility of machine learning based network intrusion detection systems, we propose to extract domain invariant features using adversarial domain adaptation from multiple network domains, and then apply an unsupervised technique for recognising abnormalities, i.e., intrusions.
More specifically, we train a domain adversarial neural network on labelled source domains, extract the domain invariant features, and train a One-Class SVM (OSVM) model to detect anomalies. At test time, we feedforward the unlabeled test data to the feature extractor network to project it into a domain invariant space, and then apply OSVM on the extracted features to achieve our final goal of detecting intrusions.
Our extensive experiments on the NIDS benchmark datasets of NFv2-CIC-2018 and NFv2-UNSW-NB15 show that our proposed setup demonstrates superior cross-domain performance in comparison to the previous approaches.
%
\end{abstract}

\end{frontmatter}

\section{Introduction}\label{introduction}

For network anomaly/intrusion detection, labelling millions of real-world network records requires a significant amount of resources and human expertise. Various attacks do not happen all the time/everywhere, and due to privacy and security concerns, labelled real-world  network intrusion detection systems (NIDS) datasets are scarce, and rarely publicly available. 
Accordingly, the common way of training machine learning (ML) based NIDSs is by using publicly available synthetic datasets.
However, as has been shown~\cite{layeghy2021benchmarking}, there is a considerable difference in the feature distribution of the benign/background traffic between real-world datasets and the synthetic benchmark datasets created in research labs.
Thus, the \textit{cross-domain} performance, i.e., the capability of correctly classifying the test samples in the presence of distribution shifts, is essential for adapting ML-based NIDSs for successful deployment and use in real-world production networks.

However, the majority of the proposed ML-based NIDSs are evaluated only on \textit{domain-specific} datasets, i.e., the training and evaluation samples are drawn from the same dataset, and cross-domain evaluation is rarely considered.

Moreover, current approaches for anomaly detection assume similar feature distributions for the training and test datasets~\cite{Erfani2016}. Therefore, these models fail to perform well when there is a distribution difference between the train (i.e., source) and test (i.e., target) data. Our study shows that the performance of the existing NIDS models degrades when they are applied in a network environment that has a different feature distribution compared to the training environment/dataset~\cite{Al-riyami2018, layeghy2022generalisability}.

To address the domain shift problem, several domain adaptation (DA) techniques have been introduced in the literature. 
These techniques try to reduce the gap between the feature representations of the labeled source and unlabeled target domains, so that the classifiers trained on the source domain perform similarly well on the target domain~\cite{Hashemi2021,Baktashmotlagh2013}. 

Generally speaking, domain adaptation approaches follow a supervised learning strategy, and thus, they are better suited to class-balanced datasets.
However, in the field of network intrusion/anomaly detection, anomalies are relatively rare events, i.e., network anomalies are in high class imbalance compared to the benign/background traffic class.
Based on our experimental results, current domain adaptation techniques based on a supervised learning strategy perform poorly on finding anomalies.

To address this gap, we are proposing a new unsupervised-learning scheme for cross-domain anomaly detection.
In this method, we use a domain adaptation technique to first extract a domain-invariant representation of the data, and then apply the anomaly detection on the projected representation.
For the domain adaptation, we use a \textit{Domain-Adversarial Neural Network (DANN)}~\cite{DANN}, which is one of the well-known and best working approaches. The DANN  can be simply incorporated in the feature extraction network by adding a gradient reversal layer to minimize the difference between the representations of the source and target domains.
We first train the DANN using the labeled source data and unlabeled target data, and then, we exploit the feature extraction branch of the DANN to obtain the domain invariant features. Finally, we apply \textit{one-class Support Vector Machines or One-Class SVM (OSVM)}~\cite{oSVM1999} on the extracted features to reach our final goal of cross-domain anomaly detection.

A network anomaly is a somewhat ill-defined concept that is used to describe any networking event such as network attacks/intrusions, failure, etc.,  that can significantly change the overall and flow-based feature statistics of a monitored network, e.g. the number of input bytes~\cite{Brauckhoff2008}.

Note that, similar to other machine learning approaches, OSVMs fails to perform well in the presence of distribution shift. Therefore, before feeding the data to an OSVM, we project the training and test data to a common subspace using a DANN.
Our results clearly indicate that projecting features to a domain invariant feature space, before feeding them into an OSVM, significantly improves the cross-domain performance of intrusion detection.

In summary, we propose a \textit{domain-invariant NIDS (DI-NIDS)} framework by leveraging recent advances in the domain adaptation literature.
DI-NIDS takes into account the intense class-imbalance nature of anomalies in the NIDS data when addressing the domain shift between the train and test datasets.
The proposed framework is evaluated against various ML-models in both  cross-domain and domain-specific evaluation scenarios, where it shows superior performance.
In the rest of this paper we first discuss the related works in the next section, then explain the proposed solution in Section~\ref{proposed method}. Extensive evaluation of DI-NIDS and comparison to the state-of-the-art are discussed in Section~\ref{Evaluation}, and Section~\ref{conclusion} concludes the paper.

\section{Related Works}\label{related works}
For the relevant related works of this paper we considered proposed NIDSs which at least follow a partial cross-domain evaluation approach across different datasets.  
We managed to find two main groups of NIDS proposals in which some aspects of the training and evaluation datasets are different.
While many of these works cannot be directly compared to our work, we still included them since they consider partial elements of cross-domain evaluation.

\subsection{Separate Source and Target Domains}
In the first group, cross-domain evaluation is realised via separate training (source domain) and test datasets (target domain).
In this group of NIDSs, techniques such as domain adaptation and transfer learning are applied on the source domain to acquire the knowledge of anomalies/attacks, and to extend this knowledge to classify samples from the target domain. 
This group can be further divided into two sub-categories; those which do not require labelled data from the target domain, and those that need a subset of the target domain to have labels.
We found three studies from the first subcategory and two studies from the second subcategory as explained below.


In the first subcategory, consisting of studies that do not need labelled data from the target domain, \cite{Zhang2020} is the only work, to the best of our knowledge, that uses a DANN in the field of network intrusion detection.
The paper focuses on detecting attacks in smart grid networks.
By applying the adversarial training to adapt learned models on normal operation data of the ISO New England grids, the authors try to detect attacks at different times of the day on their smart grid network.
The paper shows that, due to the load demand changes during different times of day, conventional ML-based NIDSs fail to detect attacks, and their proposed DANN-based method improves the detection performance. 
The authors use false data injection attacks synthesized on the IEEE 30-bus system for the evaluation of their proposed framework. The paper shows that the proposed method has superior detection performance for persistent threats recurring in a highly dynamic smart grid, compared to conventional ML-based NIDSs.

In~\cite{Fan2021} partial domain adaptation is used to map the source and target domains to a domain-invariant feature space to address the differences between the source and target datasets.
The authors use weighted adversarial networks-based domain adaptation for transferring knowledge from the publicly available labeled datasets, such as CIC-IDS2017~\cite{cic}, to an unlabelled Internet of Things (IoT) dataset, such as~\cite{kitsune2018}.
In order to evaluate their proposed framework, the authors apply it to a combination of benign traffic and various common attack classes, across two datasets. They train their model on CIC-IDS201, and evaluate it on the IoT dataset. In another evaluation, they train and test their model on different attack classes to evaluate the performance of their model for detecting unknown attacks.
The authors finally compare their framework with a DANN on a binary classification problem consisting of benign traffic and a specific attack class, and show it performs similar to a DANN.
While this is an example of a partial cross-domain evaluation of NIDSs, they do not evaluate their methods on full set of classes of the datasets. 
%
%
In addition, the evaluation only considers one direction of the cross-domain evaluation, i.e. with training on one dataset and evaluation on the other, but not vice versa. As we show in this paper, this is a significant limitation, since the results are highly asymmetric.

The authors of~\cite{Pontes2021} propose the Energy-based Flow Classifier (EFC), an anomaly-based classifier that infers a statistical model based on labelled benign samples.
They define the concept of \textit{quantum energy} for network flows and compute a threshold for the benign flows as the 95th percentile of the benign flows' energy distribution.
Then, they compare the energy of a given flow to this threshold and declare a flow as malicious if its energy is above the threshold.
The authors use three versions of the CIC-IDS~\cite{cic} dataset for the evaluation of their proposed algorithm. The paper's results are compared with conventional ML-based NIDSs for both domain-specific, i.e., the same dataset as the source and target, and cross-domain, i.e., different source and target domains.
While the reported domain-specific performance is relatively high, the cross-domain performance is significantly reduced.

While these previous studies do not require labelled data from the target domain, the next two studies need a small portion of the target domain to have labels.
The first paper in this group~\cite{Ajayi2021} considers a host-based intrusion detection approach, rather than network intrusion detection, and aims to reduce the number of labelled samples from the target domain. 
The authors use two different host-based intrusion detection datasets as the source and target domains respectively.
By using fine tuning techniques of deep learning models, such as freezing the hidden layers, they manage to reduce the number of labelled samples from the target domain, while improving the Area Under Curve (AUC) metric by 8\%.

The last paper~\cite{Singla2020} in this subcategory uses domain adaptation to address the scarcity of labeled training data by transferring the acquired knowledge from a publicly available labelled dataset. 
Initially, the authors use the UNSW-NB15~\cite{unsw} dataset and divide it into two subsets of complementary attack classes, and evaluate their approach for the same feature set assessment.
Then they use the NSL-KDD~\cite{NSL-KDD} dataset as the source and the UNSW-NB15~\cite{unsw} dataset as the target domain, and evaluate their proposed method for different feature set assessment.
Based on the provided results, the proposed method achieves a higher accuracy  for various attack classes compared to the fine tuning method, for the same number of samples from the target domain.
Although these two methods cannot be directly compared to our work, as they need labels from the target domain, they discuss the challenges of addressing the distribution gap between different intrusion detection datasets.

\subsection{Partially Different Domains}
In the second group of studies, the training and test datasets are the same, but the dataset is divided into subsets containing various attack classes.
One subset is used as the source and the other, which might include attack classes not present in the source domain, as the target domain. 
Techniques such as domain adaptation and transfer learning have been used to transfer the knowledge learnt from the attack classes available in the source domain, to classify/detect the attack classes in the target domain.

In\cite{B2020}, a transfer learning algorithm is used to transfer the knowledge of an image-based representation of network flows. The authors train a convolutional neural network (CNN) in the source domain and augment it with one dense layer in the target domain.
In ~\cite{Wu2019} the authors also use a CNN architecture for transfer learning on NIDS datasets. They use two concatenated CNNs to learn network attack patterns on a divided NSL-KDD dataset~\cite{NSL-KDD} and improve unknown/unseen attack detection performance in comparison to conventional ML-based NIDSs.
Similar to~\cite{B2020}, the authors in~\cite{fusing2020} convert divided KDD99 dataset~\cite{KDD99} records into gray-scale images which are then processed for detecting attacks using a CNN architecture.
For the purpose of transfer learning, they use samples of unseen attacks to fine tune the trained CNN.

With the exception of~\cite{Pontes2021}, none of the related works discussed in this section provides a comparable cross-domain evaluation of the proposed NIDS across different benchmark datasets. 
While the methods proposed in~\cite{Ajayi2021} and ~\cite{Singla2020} apply domain adaptation techniques on network intrusion detection datasets, they are fundamentally different to our approach, since they rely on the availability of target domain labels, which are difficult to obtain in real-world networks. 
In contrast, our method proposed in this paper does not require target domain labels, and is hence much more practical.

The last three discussed studies, i.e., \cite{B2020},~\cite{Wu2019} and~\cite{fusing2020}, do not consider domain adaptation, and mainly focus on detecting one or more attacks that were unseen during training.

While \cite{Zhang2020} and~\cite{Fan2021} are similar to our work in the sense that they do not require any target domain labels, these works do not provide a complete cross-domain evaluation such as presented in this paper. 
In~\cite{Zhang2020}, the proposed method is evaluated against the attacks injected into a smart grid network, and there is no consideration of applying such a method on the publicly available NIDS datasets.
In~\cite{Fan2021} the proposed method is evaluated against subsets of the target domain and performance metrics are provided for individual attack classes.

The method presented in ~\cite{Pontes2021} is the only approach with a cross-domain performance evaluation that is comparable to ours.  
However, the performance of the method proposed in~\cite{Pontes2021} shows a $19\%$ degradation (on average) of the cross-domain performance compared to the corresponding domain-specific performance. As we will demonstrate, our proposed method performs significantly better in this critical regard.

\section{Proposed Method (DI-NIDS)}\label{proposed method}
\begin{figure}[!b]
    \centering
    {\includegraphics[width=1\columnwidth]{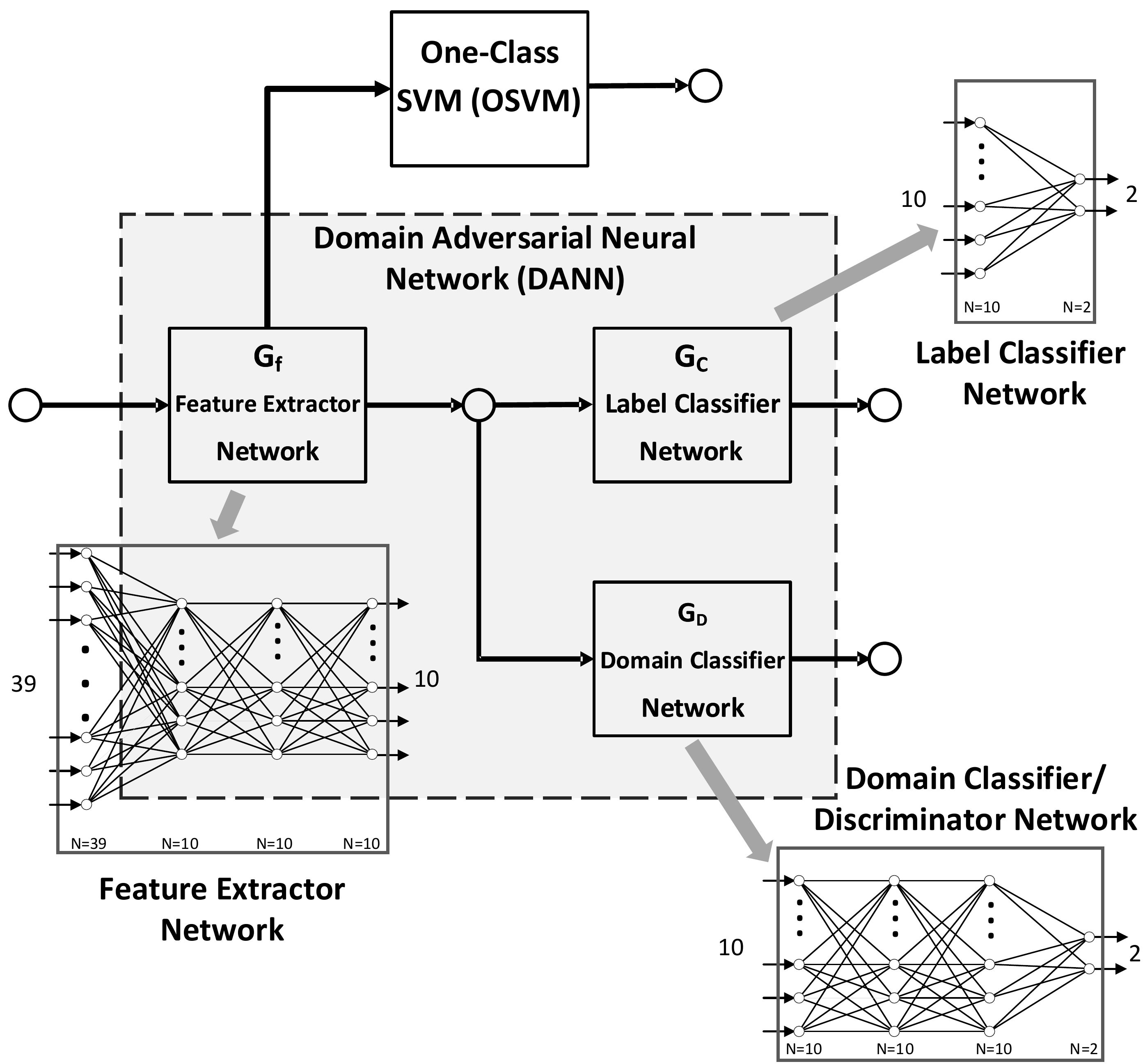}}%
    \caption{Proposed DI-NIDS architecture.}%
    \label{fig:DI-NIDS}
\end{figure}

Figure~\ref{fig:DI-NIDS} shows the architecture of \textit{Domain Invariant-Network Intrusion Detection System (DI-NIDS)}, our proposed approach, consisting of two main components, DANN and OSVM.
While OSVMs generally perform well for one-class classification problems, in particular anomaly detection, they do not perform well for cross-domain data~\cite{Oza2020} in general.
DANNs on the other hand, were designed to address the problem of domain gap, i.e., different feature distributions, in conventional machine learning models.  
%
%
%
The key idea of DI-NIDS is to enhance the domain adaptability of OSVM by leveraging the capabilities provided by a DANN.

As can be seen in Figure~\ref{fig:DI-NIDS}, a dense Multi Layer Perceptron (MLP) is used as the basis of the DANN in our model, with the  main hyper-parameters listed in Table~\ref{tab:dann-nn-params}. 
As per~\cite{DANN}, it is possible to implement a DANN using any feed-forward neural network architecture, and the type of the neural network can be selected to best match the attributes of the input data. For instance, if the input type is an image, convolutional neural networks are typically chosen. 
Since the NIDS data consist of a small number of  numeric and categorical features, we select a basic MLP network as a starting point.

%
%

\begin{table}[!b]
\footnotesize
  \centering
  \caption{Parameters of three neural networks utilised in the implemented \mbox{DI-NIDS} architecture}
    \begin{tabular}{
|>{\centering\arraybackslash}m{.6cm}
|>{\centering\arraybackslash}m{2.1cm}
|>{\centering\arraybackslash}m{.7cm}
|>{\centering\arraybackslash}m{0.8cm}
|>{\centering\arraybackslash}m{0.9cm}
|>{\centering\arraybackslash}m{0.9cm}|
}
    \hline
    Name & Function & Input Nodes & Output Nodes & No. of Hidden Layers & Hidden Layers Nodes\\
    \hline
    $G_f$ & Feature Extractor & 39 & 10 & 2 & 10\\
    \hline
    $G_C$ & Label Classifier & 10 & 2 & 0 & 0 \\
    \hline
    $G_D$ & Domain Classifier & 10 & 2 & 1 & 10\\
    \hline
    \end{tabular}%
  \label{tab:dann-nn-params}%
\end{table}%

The training of our DI-NIDS follows a two-step process.
In the first step, the DANN is trained using the source data, source labels, target data, and domain labels (labels that identify which domain the input belongs to, i.e., train or test dataset). Note that no target class labels are required in the training stage of the DANN.
Once the training stage of the DANN is completed, we employ the trained feature extractor network $G_f$ to extract domain invariant features from the data.
In the second step, the OSVM is trained using the extracted domain invariant features.
%

\noindent\textbf{DANN Component:} 
In ML-based NIDSs, as is the case in other application areas as well, it is difficult and time-consuming to label real-world data. Consequently, synthetic datasets are often used for training and evaluation of machine learning models. 
However, these synthetic datasets usually do not adequately represent real-world networks and suffer from distribution shift, i.e. a significant difference in feature distributions~\cite{layeghy2021benchmarking}. 
 Generally speaking, \textit{domain adaptation} aims at as making the distributions of source and target domains similar, so that if a classifier or detector is trained on the source data, it can perform well on the target data.~\cite{DANN}.
%

\begin{figure}[!t]
    \centering
        \subfloat
    {\includegraphics[width=0.9\columnwidth]{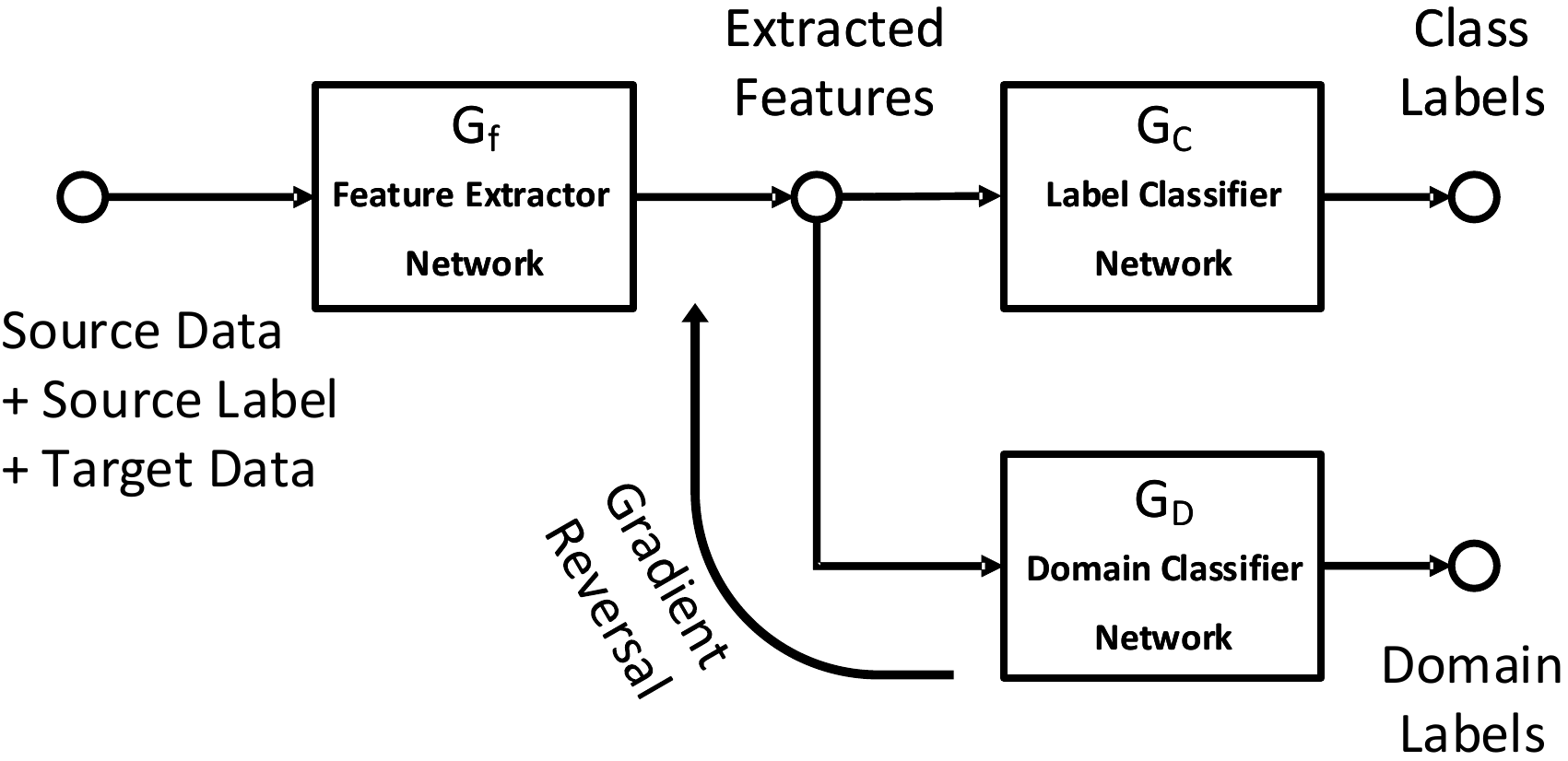}}%
    \caption{Domain Adversarial Neural Network.}%
    \label{fig:DANN}
\end{figure}

%
In this work, we employ adversarial domain adaptation to extract domain invariant features from the source and target domains. The architecture of a DANN~\cite{DANN} is shown in Figure~\ref{fig:DANN}. 
The architecture consists of three  networks: \textit{feature extractor} to extract features from the source and target domains, \textit{label classifier} to predict class labels, and \textit{domain classifier} to predict domain labels.
%
%
The domain classifier network includes a gradient reversal layer to make the distributions of the source and target features similar. Specifically, for the samples that are correctly classified by the domain classifier, a penalty is applied through multiplying their gradient by a negative factor during back propagation~\cite{Zhang2020, DANN}.

%
\textbf{\textit{Definition:}} Assume $X = \{x_1,x_2,...,x_n\}$ represents the input space, $Y=\{0,1\}$ is the set of binary labels, and $\eta: X \rightarrow Y$ is a binary classifier. 
%
%
Given the source domain, $\mathcal{D_S}=\{(x^S_{1}, y^S_{1}),(x^S_{2}, y^S_{2}),...,(x^S_{n_{S}}, y^S_{n_{1}})\}$, and the target domain, $\mathcal{D_T}=\{(x^T_{1}, y^T_{1}),(x^T_{2}, y^T_{2}),...,(x^T_{n_{T}}, y^T_{n_{2}})\}$, the feature extractor neural network $G_{f}$ can be defined as 
\begin{equation}
    G_{f}(\mathbf{x}; \mathbf{W}, \mathbf{b})=sigmoid \left(\mathbf{W}\mathbf{x} +\mathbf{b}\right)
\end{equation}
where $(\mathbf{W},\mathbf{b})$ are network weights and biases, with $x \in D_S$. The label classifier network $G_{C}$ can be written as
\begin{equation}
    G_{C}(G_{f}(\mathbf{x}); \mathbf{V}, c)=sigmoid \left(\mathbf{V}G_{f}(\mathbf{x}) +c \right)
\end{equation}
with $(\mathbf{V},c)$ representing the network parameters. Finally, the domain classifier network $G_{D}$ can be formulated as
\begin{equation}
    G_{D}(G_{f}(\mathbf{x}); \mathbf{u}, z)=sigmoid \left(\mathbf{u}^T G_{f}(\mathbf{x}) +z\right)
\end{equation}
with $(\mathbf{u},z)$ the network parameters.

More specifically, for a given sample-label pair $(x, y),\ x \in  X, y \in Y$, the domain classifier loss $L_D(x,\gamma)$ is defined as follows: 
\begin{multline}
L_D(x_i,\gamma_i) = \gamma_i log\frac{1}{G_D(G_f(\mathbf{x_{i}))}}+\\
(1-\gamma_i)log\frac{1}{1-G_D(G_f(\mathbf{x_{i}}))}
\end{multline}
where
\begin{dmath*}
\begin{cases}
\gamma_i=0 \hspace{2cm} if \hspace{.2cm} x_i \in \mathcal{D}_{S}\\
\gamma_i=1 \hspace{2cm} if \hspace{.2cm} x_i \in \mathcal{D}_{T}
\end{cases}
\label{Eq: domain-label}
\end{dmath*}
\noindent With the label classifier loss $L_y(x,y)$ defined as 
\begin{equation}
    L_y(x_{i},y_{i}) = log\frac{1}{G_{C}(G_{f}(\mathbf{x_{i}}))_{y_{i}}}
\end{equation}
the optimisation function of the domain adversarial neural network~\cite{DANN} can be written as
\begin{multline}
    \underset{\mathbf{W},\mathbf{b},\mathbf{V},\mathbf{c}, \mathbf{u},z}{min}\left[\frac{1}{n_{S}}\sum\limits_{x\in\mathcal{D}_{S}}L_y(x,y) \right.\\ 
    - \frac{\lambda}{n_{S}}\sum\limits_{x\in\mathcal{D}_{S}}L_D(x,\gamma) - 
    \left. \frac{\lambda}{n_{T}}\sum\limits_{x\in\mathcal{D}_{T}}L_D(x,\gamma)
    \right] 
\end{multline}
with $n_S$ and $n_T$ being the number of samples from the source and target domains.

\begin{figure}[!t]
    \centering
    \subfloat[\centering ][Source Label Training]
    {\includegraphics[width=0.9\columnwidth]{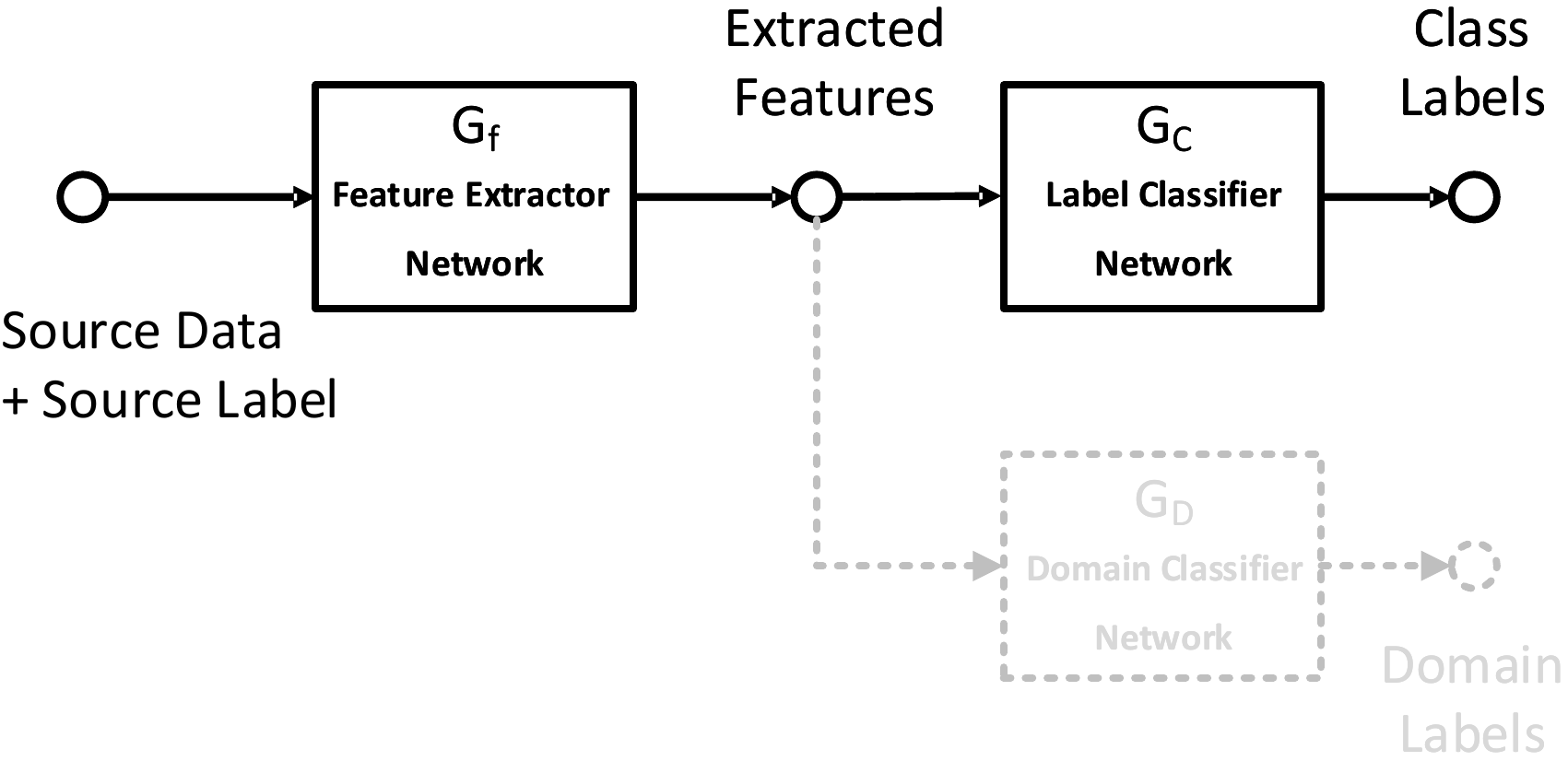}}%
    \vspace{.5cm}
    
        \subfloat[\centering ][Domain classifier training]
    {\includegraphics[width=0.9\columnwidth]{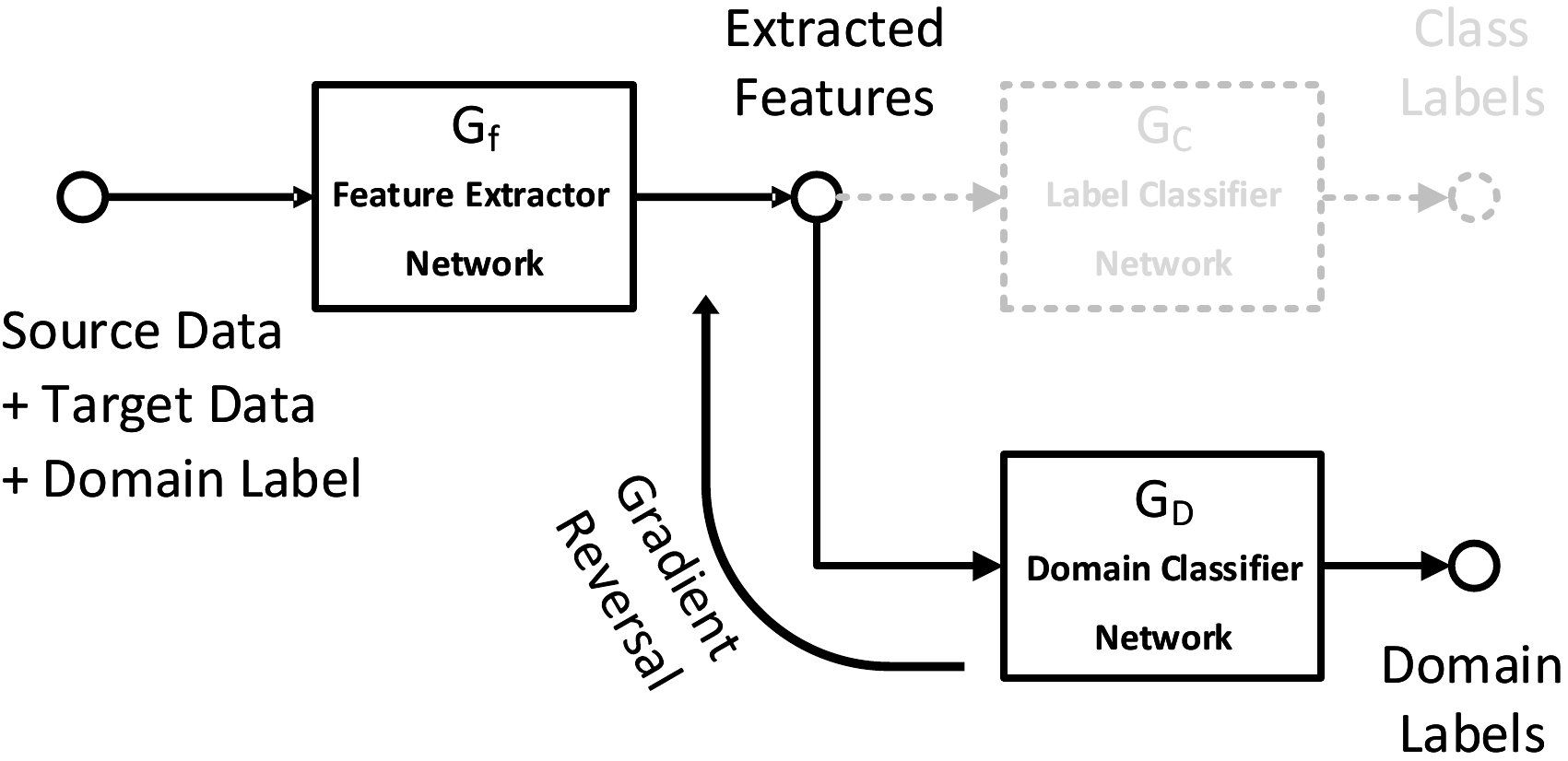}}%
    \caption{Two sub-processes of DANN including: a) Source classifier training and b) Domain classifier training}%
    \label{fig:DANN traning}
\end{figure}

Two simultaneous sub-processes of \textit{label classifier training} and \textit{domain classifier training},  as shown in Figure~\ref{fig:DANN traning}-(a) and~\ref{fig:DANN traning}-(b) respectively, contribute to the training process of the DANN.
During the label classifier training, the source data and its labels are passed through the feature extractor ($G_f$) and label classifier ($G_C$) networks, and are optimised through stochastic gradient descent to update the weights in $G_C$ and $G_f$.
%

In the domain classifier training sub-process, the source data, the target data, and \textit{domain labels} are passed through the feature extractor ($G_f$) and domain classifier ($G_D$).
The domain labels ($\gamma$) identify the domain to which a given input belongs, as defined in Equation~\ref{Eq: domain-label}.
In this sub-process, the samples with correctly predicted domain labels are penalised by the "\textit{Gradient Reversal}" layer.
The two sub-processes are optimised simultaneously, so the domain invariant and discriminative features can be learnt.

\noindent\textbf{OSVM Component:} 
In the one-class classification problem, the objective is to learn the feature distributions of the normal/benign network flows, and identify samples that deviate significantly from that distribution.
%
This is known as anomaly detection. 

\textit{Support Vector Machines (SVM)}~\cite{svm1995} were originally proposed for the multi-class classification problems and were later adopted for the one-class classification problem as proposed in~\cite{oSVM1999,oSVM_int_2013}.

%
Assume $\chi = \{(x_1, y_1),(x_2,y_2),...,(x_n,y_n)\}$ represents a dataset in which  $x_i \in \mathbb{R}^d$ is the $i$th data point in a $d$-dimensional input space $\mathbb{I}$, and $y_i\in \{-1,1\}$ represents the $i$th output, i.e., class labels.
A SVM uses a nonlinear function $\phi$, i.e., a \textit{kernel}, to project data points from their input space $\mathbb{I}$ to a high dimensional feature space $\mathbb{F}$, in which the classes can be distinguished by a linear hyperplane $w^Tx+b=0$, with $w\in\mathbb{F}$ and $b\in\mathbb{R}$. To find this hyperplane, SVM minimises the following objective function~\cite{oSVM_int_2013}:
\begin{equation}
    \underset{w,b,\xi_{i}}{min}\left[\frac{\left\| w \right\|^2}{2}+C\sum\limits_{i=1}^{n}\xi_{i}\right]
\end{equation}
subject to the two constraints of:
\begin{dmath*}
\begin{cases}
y_i\left( w^t \phi \left( x_i \right) + b \right) \ge 1 - \xi_{i} \hspace{1cm} \forall\ i=1,...,n \\ \\
\xi_{i} \ge 0 \hspace{3.9cm} \forall\ i=1,...,n
\end{cases}
\end{dmath*}
Here, $C$ is a constant determining the number of training data points within the margin between two classes, i.e., training error, and $\xi$ is the \textit{slack variable} to prevent overfitting. Solving this minimisation problem via Lagrange multipliers results in the following classification rule for a given data point $x$~\cite{oSVM_int_2013}:

\begin{equation}
    f(x)=sgn\left[\sum\limits_{i=1}^{n}\alpha_{i} y_i K\left(x,x_i\right)+b\right]
\end{equation}
with $\alpha_{i}$ being the Lagrange multipliers and the function $K(x,x_i)={\phi(x)}^T{\phi(x)}$ being the \textit{kernel function}.

In OSVM~\cite{oSVM1999}, instead of learning a hyperplane to separate two classes of data, a hyperplane is learnt to separate the abnormal data points from the normal density in the origin. Hence, the intention is to maximise the distance of the learnt hyperplane from the origin in the feature space $\mathbb{F}$.
An OSVM can be mathematically formulated as a minimisation of the below objective function~\cite{oSVM_int_2013}:
\begin{equation}
    \underset{w,\xi_{i},\rho}{min}\left[\frac{\left\| w \right\|^2}{2}+\frac{1}{\nu n}\sum\limits_{i=1}^{n}\xi_{i}-\rho\right]
\end{equation}
subject to the following two constraints:
\begin{dmath*}
\begin{cases}
\left( w . \phi \left( x_i \right) \right) \ge \rho - \xi_{i} \hspace{2cm} \forall\ i=1,...,n \\ \\
\xi_{i} \ge 0 \hspace{3.9cm} \forall\ i=1,...,n
\end{cases}
\end{dmath*}
with $\nu \in \left(0,1\right)$ being the upper bound identifier for the fraction of outliers and lower bound on the support vectors, and $\rho \in \mathbb{R}$ being the offset.
The Lagrange method is used to solve the above minimisation problem which results in the following classification rule~\cite{oSVM_int_2013}:
\begin{multline}
    f(x)=sgn\left( w \phi \left( x_i \right) - \rho\right)\\
     =sgn\left[\sum\limits_{i=1}^{n}\alpha_{i} y_i K\left(x,x_i\right)-\rho\right]
\end{multline}
The hyperplane identified by $w$ and $\rho$ has the maximum distance to the origin in the feature space $\mathbb{F}$, which separates anomalous data points from the normal ones concentrated in the origin.

%
%
%

\section{Experimental Evaluation}\label{Evaluation}

\subsection{Datasets}\label{Datasets}
For the evaluation of our proposed approach and models, we used the NetFlow versions of two publicly available NIDS datasets, UNSW-NB15~\cite{unsw} and CIC-2018~\cite{cic}.
These are the two most-cited NIDS benchmark datasets among the recent NIDS datasets, providing a  more realistic representation of today's  network traffic in comparison to older benchmark datasets such as NSL-KDD~\cite{NSL-KDD}.

The original UNSW-NB15 dataset was generated and published by researchers at the University of New South Wales at the Australian Defence Force Academy Canberra in 2015. 
The second dataset, CIC-2018, was collected from a completely different network setup and published by the Canadian Institute for Cybersecurity (CIC), University of New Brunswick (UNB), in 2018.

The original versions of these datasets are published with two very different feature sets, with only six common features across the two sets, out of a total of 42 and 75 features of UNSW-NB15 and CIC-2018 datasets respectively~\cite{sarhan2020netflow}. 
This generally makes it impossible to fairly compare the performance of ML models on both datasets.

Therefore, a common feature set was needed for the cross-domain evaluation of our ML models. 
Accordingly, we used the NFv2-UNSW-NB15 and NFv2-CIC-2018 datasets, that were converted to NetFlow from their original formats in ~\cite{sarhan2021towards}.
NetFlow is the \textit{de facto} standard in network flow reporting and is widely deployed in real-world networks.
The features in the NetFlow versions of these datasets are comprised of 43 NetFlow version 9 fields, which represent bi-directional network flows\footnote{A previous version of these datasets with 20 NetFlow fields is also publicly available~\cite{sarhan2020netflow}.}.

%
%

%
%
%
%

\begin{figure}[!t]
    \centering
    \subfloat[\centering ][NFv2-UNSW-NB15]
    {\includegraphics[width=0.75\columnwidth]{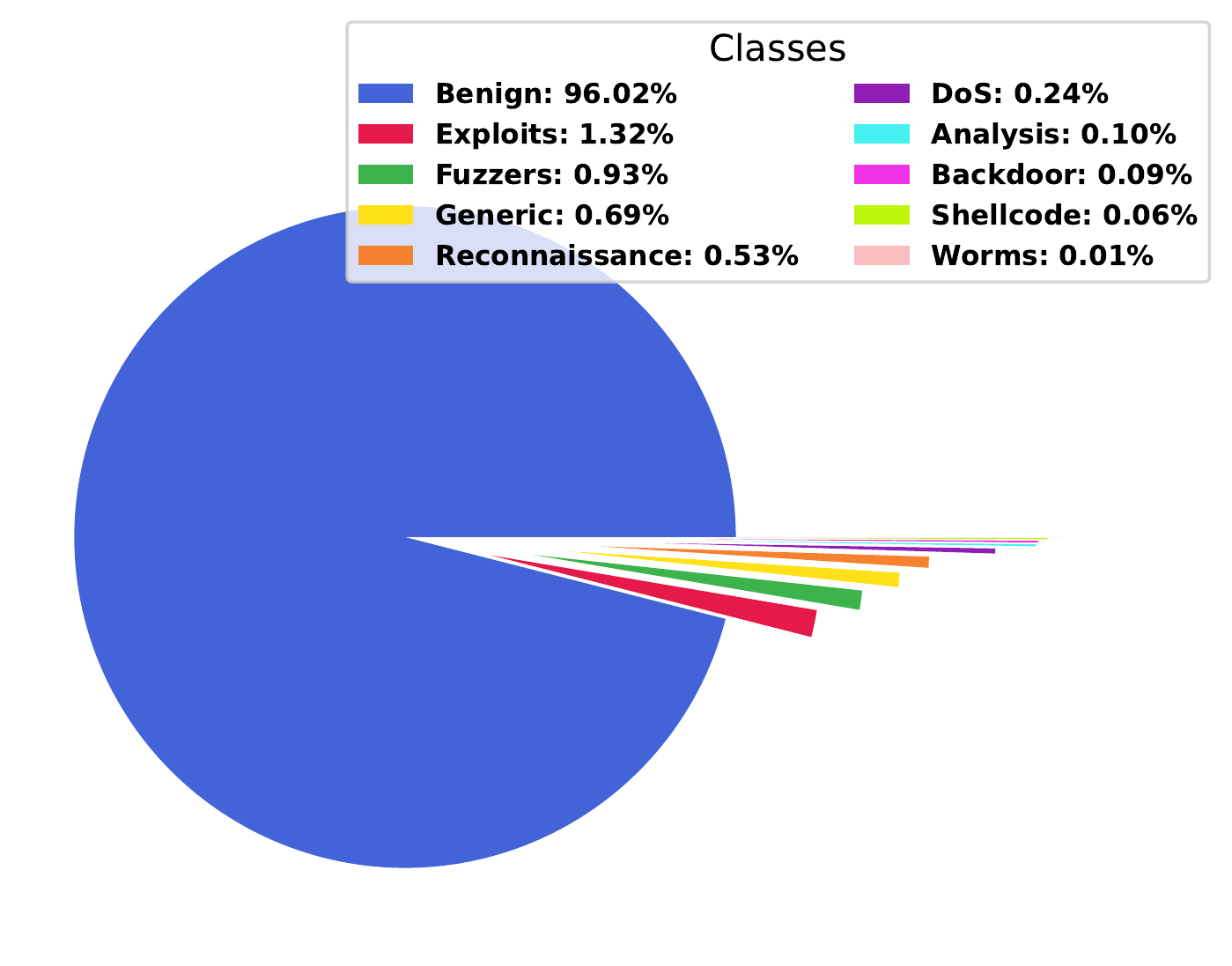}}%
    \vspace{0.1cm}
    \subfloat[\centering ][NFv2-CIC-2018]
    {\includegraphics[width=0.8\columnwidth]{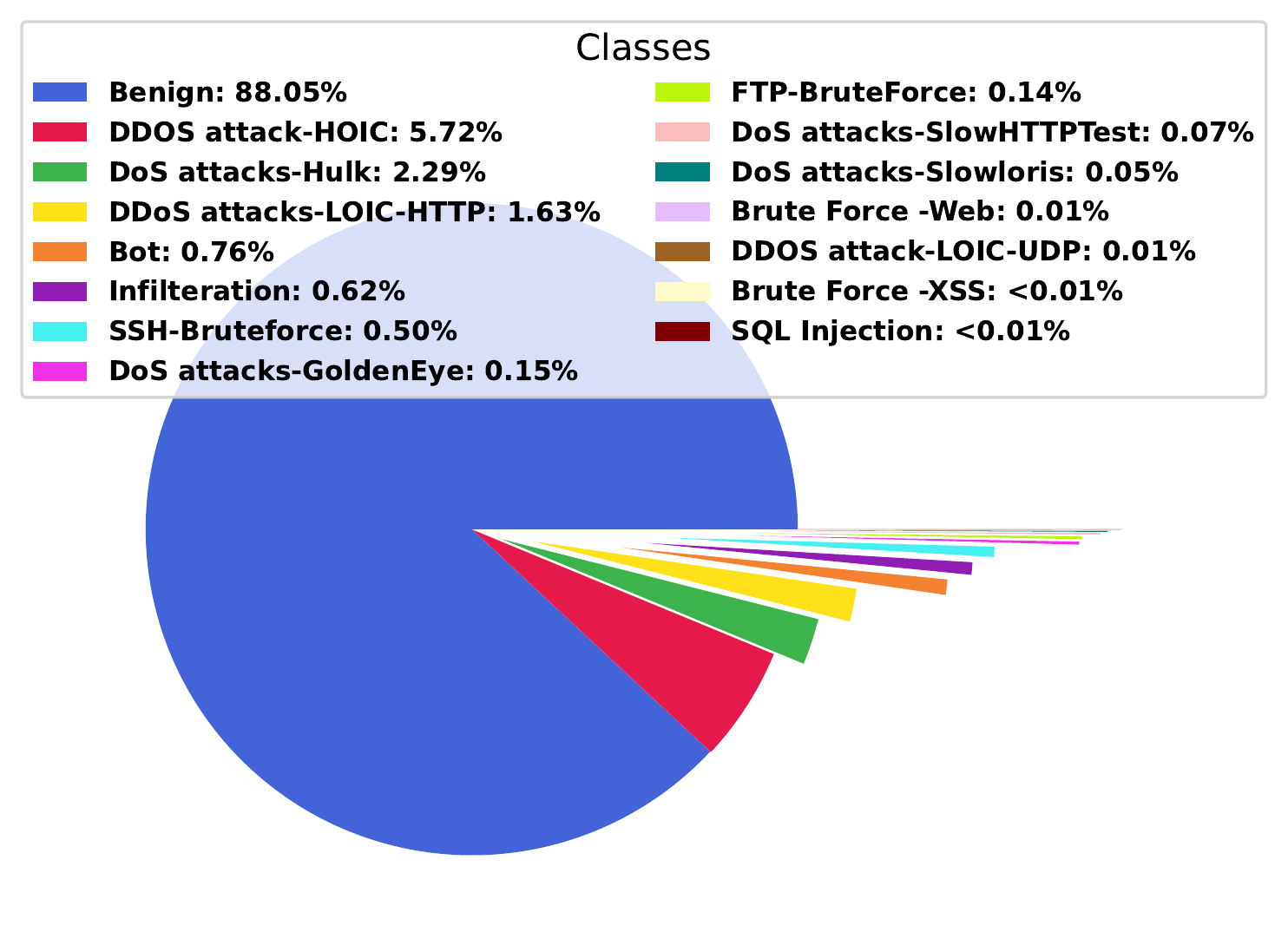}}%
    \caption{Class distribution for the datasets used in this study  a)NFv2-UNSW-NB15 and b)NFv2-CIC-2018}%
    \label{fig:NF_meta_piechart}
\end{figure}

%
Figure~\ref{fig:NF_meta_piechart} shows the class distributions for these two datasets.
The NFv2-UNSW-NB15 dataset includes 2,390,275 flows, labelled as either Benign/background traffic, or one of the 9 attack classes. The Benign class makes up 96.02\% of the entire dataset.
The NFv2-CIC-2018 dataset consists of 18,893,708 flows, which are either Benign or belong to one of the 14 attack classes, including various DoS and DDoS attacks, SQL Injection, Infiltration and Brute Force attacks. In this dataset, 88.05\% of the flows are Benign.

Since the focus of this paper is on binary classification,  all the various attack classes in each dataset are aggregated into a single class called \textbf{Attack}. 
However, the nature of this Attack class is totally different for each dataset. Indeed, the number of attack types, the number of flows belonging to each attack type and the ratio of number of flows in each attack type to total flows are different in each dataset. Even the attacks with  similar names from two datasets, such as \textit{DoS} (from NFv2-UNSW-NB15) and \textit{DoS attacks-Slowloris} (from NFv2-CIC-2018) represent completely different types of attacks.

Consequently, these datasets not only represent different network environments, as they have been generated in completely different networks, they represent domains with different label sets.
This is also reflected in previous studies such as~\cite{layeghy2022generalisability}, where it is shown that the performance of conventional ML models trained on one of theses datasets severely degrades when tested on the other dataset.
Another previous study~\cite{layeghy2021benchmarking} has also shown the difference between the feature distributions in the benign/background traffic of various NIDS datasets.

\begin{figure}[!t]
    \centering
    \subfloat[\centering ][Embedded Input Data]
    {\includegraphics[width=0.85\columnwidth]{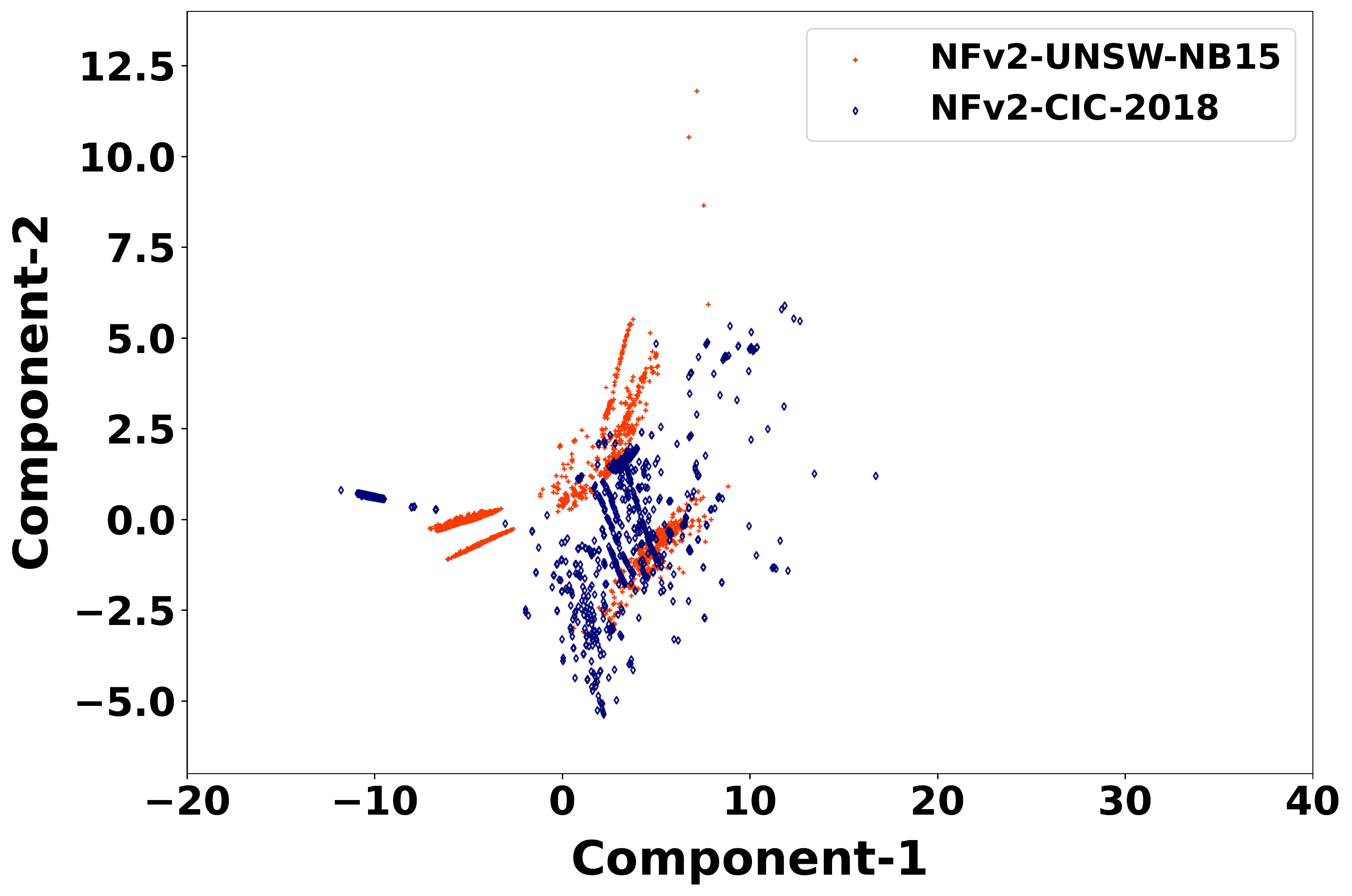}}%
    \vspace{0.1cm}
    \subfloat[\centering ][Embedded DANN-1 Projected Data]
    {\includegraphics[width=0.85\columnwidth]{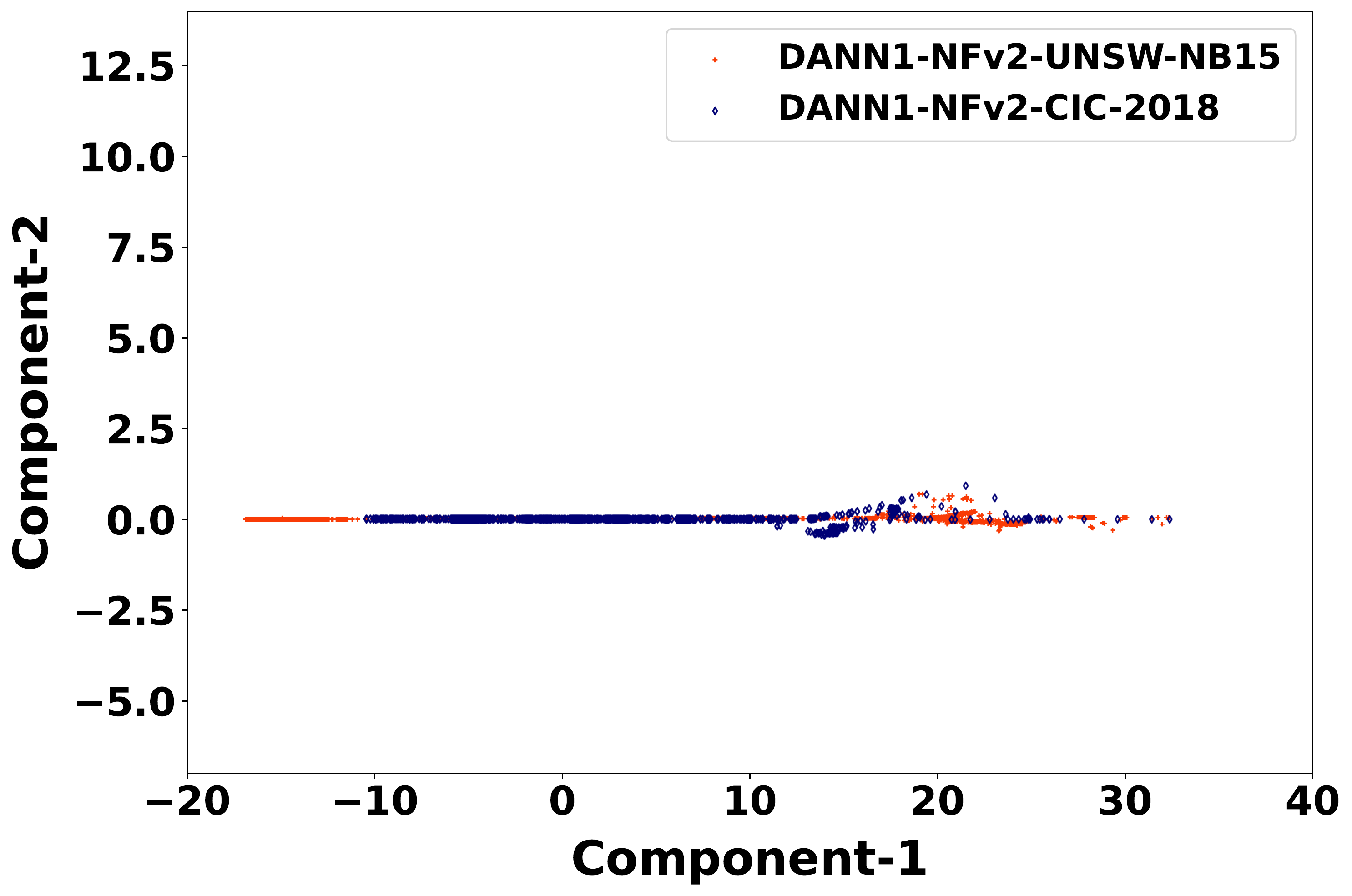}}%
    \vspace{0.1cm}
        \subfloat[\centering ][Embedded DANN-2 Projected Data]
    {\includegraphics[width=0.85\columnwidth]{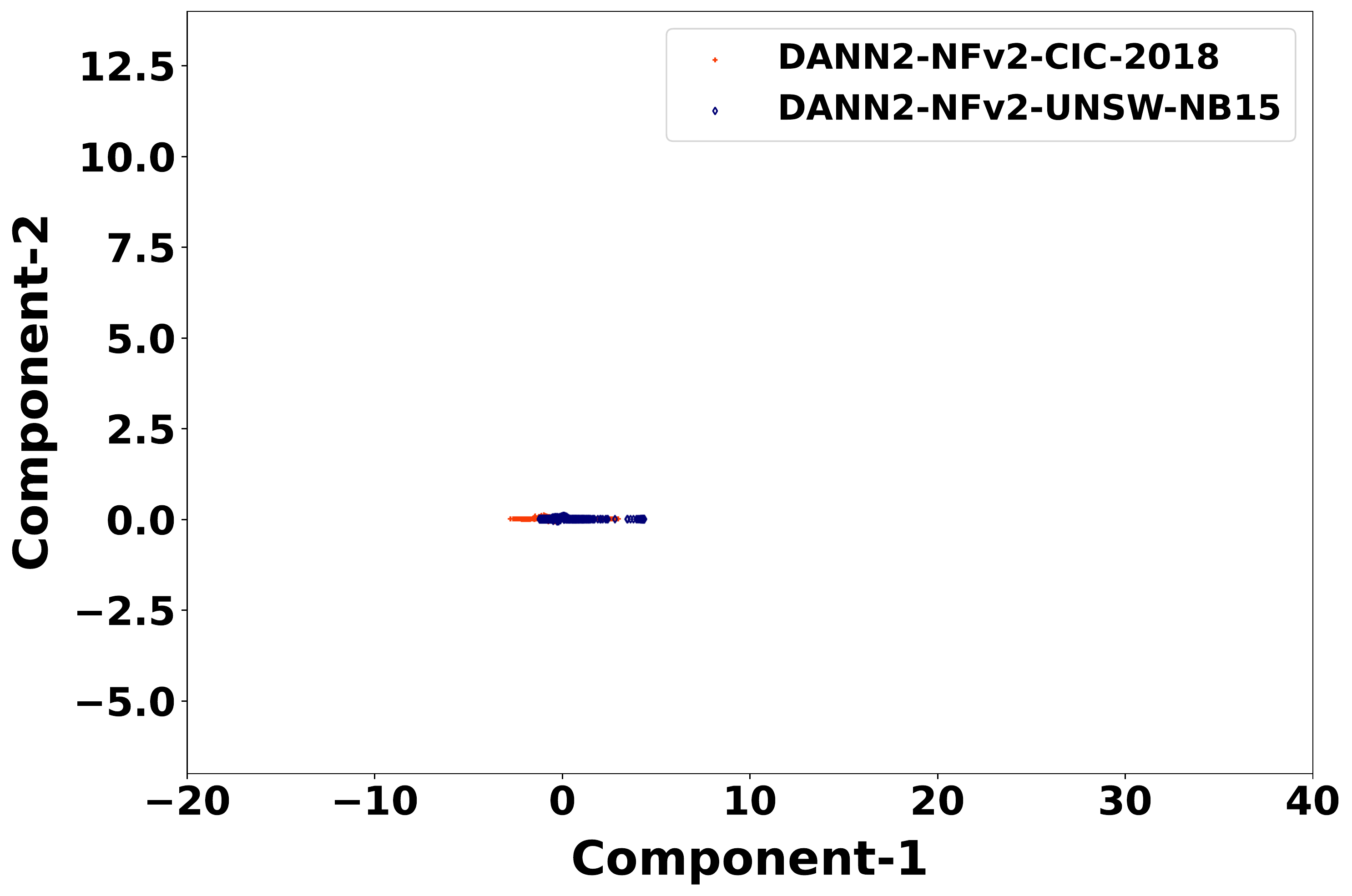}}%
    \vspace{0.5cm}
    \caption{Illustrating the domain shift  before and after DANN projection for the two NIDS datasets used in this study. (a) Input data, and DANN projection via training on b) NFv2-UNSW-NB15 and (c) NFv2-CIC-2018}%
    \label{fig:emb-1}
\end{figure}

\subsection{\textbf{Domain Invariant Projection}}
The main purpose of the DANN component is the projection of the input data into a domain invariant space. 
In order to investigate this for the two NIDS datasets used in this study, we compare the separation of the two datasets before and after the domain invariant projection via DANN.
%
%
The input datasets have 43 features, and after passing them through the Feature Extractor ($G_f$) of the DANN component,  are reduced to 10 features (number of nodes on the output layer of $G_f$), as can be seen in Figure~\ref{fig:DI-NIDS}.

Figure~\ref{fig:emb-1}-(a) shows the visualisation of the two \textit{embedded} input datasets. 
This visualisation is generated by reducing the dimensionality of the data from 43 to 2, using the isomap~\cite{isomap} embedding algorithm.
Similarly,  Figure~\ref{fig:emb-1}-(b) and (c) show the visualisation of datasets after projection via DANN ($G_f$ network), trained on NFv2-UNSW-NB15 (referred to as DANN1) and trained on NFv2-CIC-2018 (referred to as DANN2) datasets respectively. 
As in the case of the original input datasets, isomap is used to reduce the dimensions of the DANN-projected datasets from 10 to 2.
As can be seen, the separation of the two embedded input datasets is significant in Figure~\ref{fig:emb-1}-(a). This means that there is a considerable distribution shift between the two input datasets.
Accordingly, it is not expected that an ML model trained on one of these datasets to perform well when evaluated on the other dataset. 
However, after the DANN projection, as can be seen in both Figure~\ref{fig:emb-1}-(b) and Figure~\ref{fig:emb-1}-(c), the embedded data points show significantly less separation, indicating that the feature distributions are much closer, which was the aim.
Hence, we expect an ML model, such as OSVM, trained on either of the DANN projected features, to perform well on the other one, and we therefore expect the DI-NIDS to exhibit a higher degree of domain invariance. 
We will confirm this expectation via a detailed experimental evaluation of DI-NIDS in following subsection. 



\begin{table}[!b]
  \centering
  \caption{The model parameters for two deep learning-based NIDSs along with the other parameters for training and evaluation of the models}
  \begin{tabular}{|c|c|c|}
    \hline
      \textbf{Parameter}    & \textbf{Feed Forward} & \textbf{LSTM} \\[0.5ex]
              \hline
    \textbf{No. Hidden Layers} & 3     & 4 \\[0.5ex]
    \hline
    \textbf{No. Nodes (each layer)} & 10    & 10 \\[0.5ex]
    \hline
    \textbf{Learning Rate} & 0.0001 & 0.0001 \\[0.5ex]
        \hline
    \textbf{Dropout Ratio} & 0.2   & 0.2 \\[0.5ex]
    \hline
    \textbf{Batch Size} & 512   & 512 \\[0.5ex]
    \hline
    \textbf{Validation Split} & 0.3   & 0.3 \\[0.5ex]
    \hline
    \textbf{No. of Folds} & 5     & 5 \\[0.5ex]
    \hline
    \end{tabular}%
  \label{tab:nn_params}%
\end{table}%

\begin{table}[!t]
  \centering
  \caption{The model parameters for two shallow learning-based NIDSs along with the other parameters for training and evaluation of the models}
    \begin{tabular}{|c|c|c|}
    \hline
     Parameter     & \textbf{Random Forest} & \textbf{Extra Tree} \\[0.5ex]
    \hline
    \textbf{ccp\_alpha} & 0.001 & 0.001 \\[0.5ex]
    \hline
    \textbf{Batch Size} & 512   & 512 \\[0.5ex]
    \hline
    \textbf{Validation Split} & 0.3   & 0.3 \\[0.5ex]
    \hline
    \textbf{No. of Folds} & 5     & 5 \\[0.5ex]
    \hline
    \end{tabular}%
  \label{tab:shallow_params}%
\end{table}%

\subsection{\textbf{Results}}\label{results}
In order to evaluate the proposed DI-NIDS framework, we performed three sets of experiments.
In the first set of experiments, DI-NIDS was evaluated in a domain-specific setup on our chosen two benchmark  datasets, i.e., it was trained and tested on the same dataset, for each dataset separately.
In the second and third sets of experiments, we evaluated DI-NIDS in a cross-domain setup, i.e., with training on one dataset and testing on the other. 
First, one dataset was used as the train/source, and the other dataset as the test/target dataset. 
Then, we swapped the train/source and test/target datasets and ran the evaluation again, in order to obtain the cross-domain performance in both directions.
The domain-specific evaluation mainly serves as a baseline, i.e., to compare the performance to the cross-domain evaluation scenario.

Since there is no previous study of domain adaptation using the same NIDS datasets that we are using in this study, we rely on the results published in one of our previous works ~\cite{layeghy2022generalisability}, which includes both cross-domain and domain-specific evaluation of conventional deep and shallow machine learning models.
The models used from this study include a long short-term memory (LSTM) model and two shallow learning methods, i.e.,  Extra-Tree and Random-Forest.
In addition to the model performance results provided in~\cite{layeghy2022generalisability}, we also used an MLP (Feed Forward) model in this study, to provide an additional baseline result.  
This MLP model is similar to the MLP model utilised to create the DANN, i.e., a combination of  the $G_f$ and $G_C$ blocks of the DANN, as shown in  Figure~\ref{fig:DANN traning}-(a). 
Selecting similar models in the DANN component of \mbox{DI-NIDS} and the baseline MLP allowed us to evaluate the role of the augmenting block $G_D$ in the cross-domain evaluation.
Tables~\ref{tab:nn_params} and~\ref{tab:shallow_params} show the parameters of the conventional deep and shallow learning models used for the comparison.

In addition to these conventional ML models, we also included the OSVM and DANN blocks individually in our evaluation.
This allowed us to evaluate the performance of DI-NIDS and compare it with the performance of its key building blocks separately. 
For the evaluation of OSVM and DANN, each of these models was separately trained and evaluated. 
In the case of domain-specific evaluation, these models were trained and tested on NFv2-UNSW-NB15 and NFv2-CIC-2018 datasets separately.
In the case of the cross-domain evaluation, each model was once trained on NFv2-UNSW-NB15 and tested on  NFv2-CIC-2018 and then trained on NFv2-CIC-2018 and tested against NFv2-UNSW-NB15.

\begin{figure}[b]
    \centering
    {\includegraphics[width=1\columnwidth]{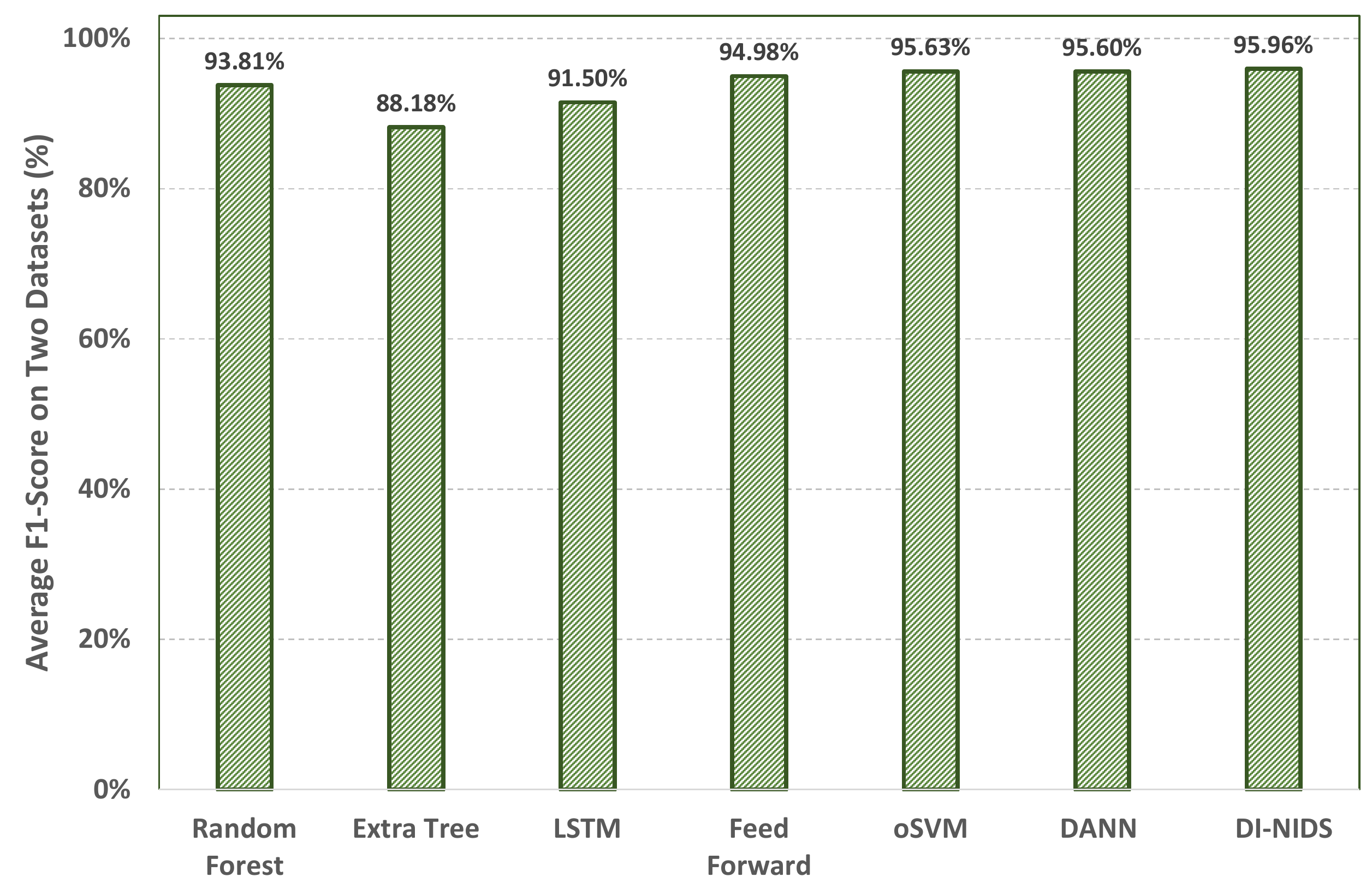}}%
    \caption{Average domain-specific F1-Score of various ML models on two datasets NFv2-CIC-2018 and NFv2-UNSW-NB15 (shown in Table~\ref{tab:self-eval-f1}) where the test and training datasets are the same}%
    \label{fig:ds-avg-f1-score}
\end{figure}

Table~\ref{tab:self-eval-f1} shows the results (F1-Score) of domain-specific performance evaluation for DI-NIDS, and the six baseline ML models used for comparison. 
As can be seen, while the performance of all the models are largely similar on both NIDS datasets, DANN and DI-NIDS show the best performance on the NFv2-CIC-2018 and  NFv2-UNSW-NB15 datasets respectively.

Figure~\ref{fig:ds-avg-f1-score} shows the average domain-specific performance (F1-Score) of the considered models for the two datasets, as stated in Table~\ref{tab:self-eval-f1}.
As can be seen, DI-NIDS has the highest average performance on these two datasets, closely followed by OSVM and DANN.
While the main idea of proposing DI-NIDS is to address the low  cross-domain performance of conventional ML-based NIDSs, the fact that DI-NIDS has the highest domain-specific performance among all the considered models is very promising.

\begin{table}[!t]
  \centering
  \caption{Domain-specific performance (F1-Score (\%))  of various ML models compared to DANN, OSVM and DI-NIDS when trained and evaluated on the same dataset}
    \begin{tabular}
{
|>{\centering\arraybackslash}m{2.8cm}
|>{\centering\arraybackslash}m{2cm}
|>{\centering\arraybackslash}m{2.4cm}|
}

\hline
\textbf{ML Model} & \textbf{NFv2-CIC-2018} & \textbf{NFv2-UNSW-NB15} \\
    \hline
    Random-Forest~\cite{layeghy2022generalisability} & 95.44\% & 92.17\% \\
    Extra Tree~\cite{layeghy2022generalisability} &  84.62\% & 91.73\% \\
    LSTM NN~\cite{layeghy2022generalisability} & 90.17\% & 92.82\% \\
    Feed Forward NN$^*$ & 97.72\% & 92.24\% \\
    OSVM & 92.97\% & 98.28\% \\
    \textbf{DANN} & \textbf{97.81}\% & 93.38\% \\
        \hline
    \textbf{DI-NIDS} & 93.23\% & \textbf{98.68}\% \\
    \hline
    \multicolumn{3}{c}{* The Feed Forward Neural Network as depicted in Figure~\ref{fig:DANN traning}-a}\\
    \end{tabular}%
  \label{tab:self-eval-f1}%
\end{table}%

In the next set of experiments, we compared the performance of DI-NIDS to the other models in two cross-domain setups. 
In this setting, first each model is trained on the NFv2-CIC-2018 dataset, and then tested against the NFv2-UNSW-NB15 dataset.Then the source and target domains are swapped, i.e., the models are trained on NFv2-UNSW-NB15 and tested against NFv2-CIC-2018.
Tables~\ref{tab:cd-source-cic} and \ref{tab:cd-source-unsw} show the results of these evaluations respectively.
Similar to the domain-specific evaluation, we used the results of the Random-Forest, Extra-Tree and LSTM models on the same datasets from~\cite{layeghy2022generalisability} 
for cross-domain evaluation.
For the Feed Forward model (MLP), similar to the domain-specific evaluation, we used the same network setting/parameters as mentioned in Table~\ref{tab:nn_params}.

\begin{table}[!b]
  \centering
  \caption{Cross-domain performance (F1-Score (\%)) and its difference to the corresponding domain-specific performance of DI-NIDS and conventional ML models for the case where source domain is \textbf{NFv2-CIC-2018} and the target domain is \textbf{NFv2-UNSW-NB15}}
    \begin{tabular}
{
|>{\centering\arraybackslash}m{2.9cm}
|>{\centering\arraybackslash}m{2.4cm}
|>{\centering\arraybackslash}m{2.2cm}|
}

\hline
\textbf{ML Model} & \textbf{F1-Score (\%)} & \textbf{Performance Degradation} \\
    \hline
    Random-Forest~\cite{layeghy2022generalisability} & 0.84\% & 94.60\% \\
    Extra Tree~\cite{layeghy2022generalisability} & 0.57\% & 84.05\% \\
    LSTM NN~\cite{layeghy2022generalisability} & 9.63\% & 80.54\% \\
    Feed Forward NN$^*$ & 3.09\% & 94.63\% \\    
    \textbf{OSVM} & \textbf{86.15}\% & \textbf{6.79}\% \\
    DANN & 17.31\% & 80.50\% \\
        \hline
    DI-NIDS & 85.79\% & 7.44\% \\
    \hline
    \multicolumn{3}{c}{* The Feed Forward Neural Network as depicted in Figure~\ref{fig:DANN traning}-a}
    \end{tabular}%
  \label{tab:cd-source-cic}%
\end{table}%

\begin{table}[!t]
  \centering
  \caption{Cross-domain performance (F1-Score (\%)) and its difference to the corresponding domain-specific performance of DI-NIDS and conventional ML models for the case where source domain is \textbf{NFv2-UNSW-NB15} and the target domain is \textbf{NFv2-CIC-2018}}
    \begin{tabular}
{
|>{\centering\arraybackslash}m{3cm}
|>{\centering\arraybackslash}m{2.3cm}
|>{\centering\arraybackslash}m{2.2cm}|
}

\hline
\textbf{ML Model} & \textbf{F1-Score (\%)} & \textbf{Performance Degradation} \\
    \hline
    Random-Forest~\cite{layeghy2022generalisability} & 7.70\% & 84.47\% \\
    Extra Tree~\cite{layeghy2022generalisability} &  17.47\% & 74.26\% \\
    LSTM NN~\cite{layeghy2022generalisability} & 14.20\% & 78.62\% \\
    Feed Forward NN$^*$ & 30.79\% & 61.45\% \\    

    OSVM & 15.74\% & 82.54\% \\
    DANN & 61.94\% & 31.44\% \\
        \hline
    \textbf{DI-NIDS} & \textbf{93.29}\% & \textbf{5.39}\% \\
    \hline
    \multicolumn{3}{c}{* The Feed Forward Neural Network as depicted in Figure~\ref{fig:DANN traning}-a}\\
    \end{tabular}%
  \label{tab:cd-source-unsw}%
\end{table}%

In both tables, the first column shows the model name, the second column shows its cross-domain performance and the third column shows the difference between the cross-domain performance and its corresponding domain-specific evaluation.
For instance, in Table~\ref{tab:cd-source-cic} the Random-Forest model has a F1-Score of $0.84\%$ when trained on NFv2-CIC-2018 and tested against NFv2-UNSW-NB15.
This is $94.60\%$ lower than its performance (F1-Score) when trained and tested on NFv2-CIC-2018, as shown in the first column of Table~\ref{tab:self-eval-f1}.
Accordingly, the third column of Table~\ref{tab:cd-source-cic} shows $94.60\%$ for the performance degradation of the Random-Forest model.

As can be seen, OSVM is the best performing model when NFv2-CIC-2018 is the source and NFv2-UNSW-NB15 is the target domain. 
In this case, the performance (F1-Score) of DI-NIDS is only $0.36\%$ lower than ons-class SVM, i.e., $85.79\%$.
While this is a great result for OSVM in regards to domain adaptation, the next evaluation shows an entirely different results.

The results in Table~\ref{tab:cd-source-unsw} indicate that DI-NIDS is the best performing model in the second experiment, with a significant advantage over the other models.
In fact, DI-NIDS is the only model capable of maintaining its performance over the two cross-domain experiments.
It has a $7.44\%$ performance degradation in the first cross-domain experiment and $5.39\%$.
OSVM, which was the best performing model in the first experiment, only achieves an F1-Score of $15.74\%$ in this experiment, with $82.54\%$ degradation compared to its corresponding domain-specific performance.

\begin{figure}[!b]
    \centering
    {\includegraphics[width=1\columnwidth]{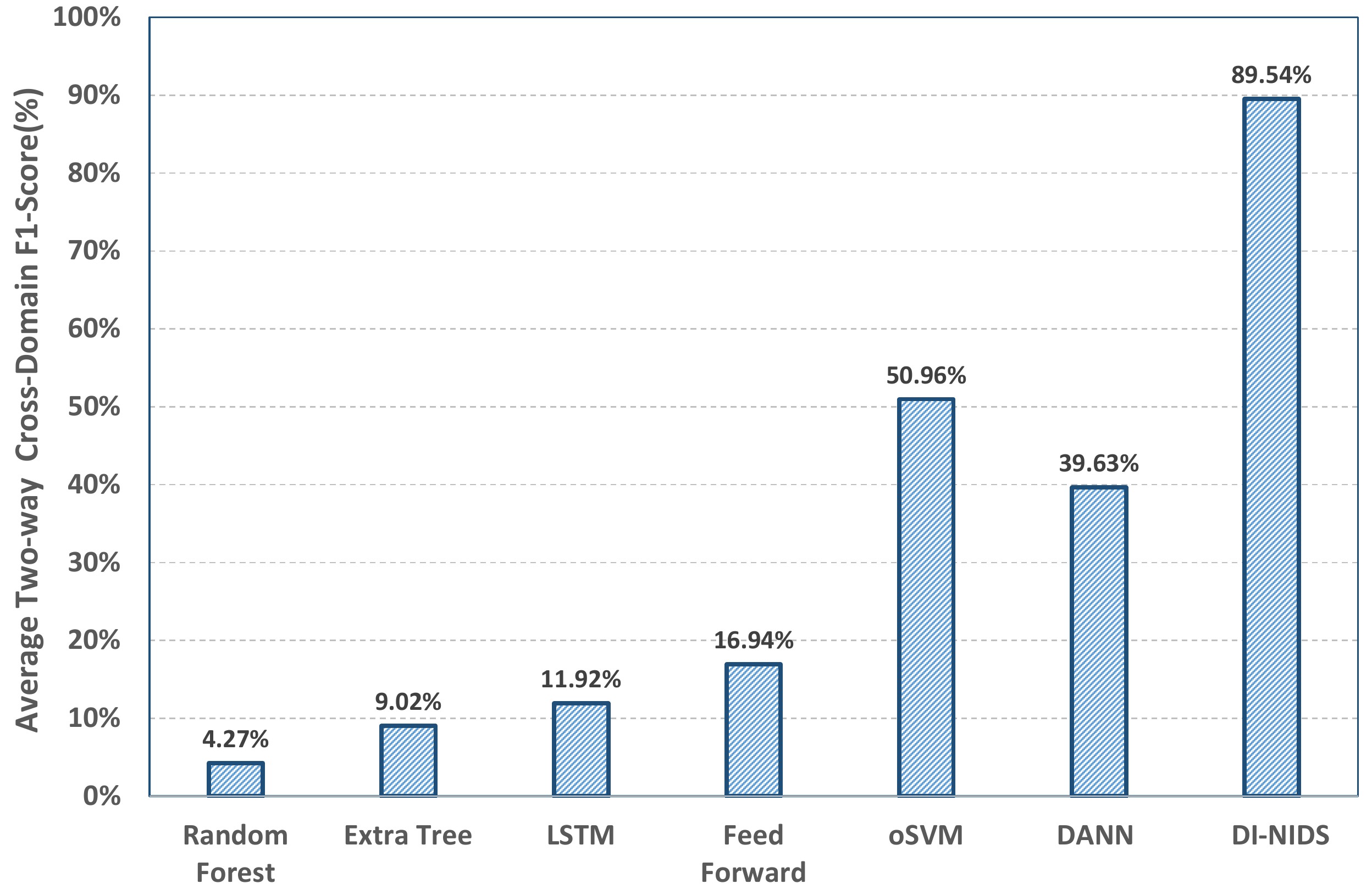}}%
    \caption{Averaged cross-domain performance (F1-Score) of ML models across two cross-domain experiments}%
    \label{fig:two-way-drop}
\end{figure}

Figure~\ref{fig:two-way-drop} shows the average performance (F1-Score) of the models for the two cross-domain experiments.
As can be seen, all the conventional ML models, including Random-Forest, Extra-Tree, LSTM and Feed Forward, have very low average cross-domain performance, no better than $16.94\%$.
The OSVM and DANN models are somewhat better, with average cross-domain performances of $50.96\%$ and $39.63\%$ respectively.
However, they are still far from a truly domain invariant model, i.e., a model that is capable of maintaining its performance in the presence of domain shifts. This is in clear contrast to DI-NIDS, which largely maintains its performance, with an average  cross-domain performance of 89.54/

\begin{table}[!b]
\scriptsize
  \centering
  \caption{Cross-domain performance comparison degradation (Degr.) comparison}
    \begin{tabular}
    {
|>{\centering\arraybackslash}m{0.5cm}
|>{\centering\arraybackslash}m{1.6cm}
|>{\centering\arraybackslash}m{1.5cm}
|>{\centering\arraybackslash}m{1cm}
|>{\centering\arraybackslash}m{0.7cm}
|>{\centering\arraybackslash}m{0.7cm}|
}
\hline    \multicolumn{1}{|c|}{\textbf{Model}} & \textbf{Train Dataset} & \textbf{Test Dataset} & \textbf{F1-Score (\%)} & \textbf{Degr. (\%)} & \textbf{Avg. Degr. (\%)} \\
    
    \hline
    \hline
    \multirow{4}[8]{*}{\rotatebox{90}{\textbf{Pontes et al. \cite{Pontes2021} }}} & \multirow{2}[4]{*}{{CIC-2017 }} & \vspace{0.2cm}{CIC-2017} & \vspace{0.2cm}89.80  & \multirow{2}[4]{*}{\textbf{11.10}} & \multirow{4}[8]{*}{\textbf{19.30}} \\[2ex]
    
\cline{3-4}   &       & \vspace{0.2cm}\cellcolor{gray!25}{CIC-2019} & \vspace{0.2cm}\cellcolor{gray!25}78.70  &       &  \\ [2ex]
\cline{2-5}          & \multirow{2}[4]{*}{{CIC-2019}} & \vspace{0.2cm}{CIC-2019} & \vspace{0.2cm}91.60  & \multirow{2}[4]{*}{\textbf{27.50}} &  \\ [2ex]
\cline{3-4}          &       & \vspace{0.2cm}\cellcolor{gray!25}{CIC-2017} & \vspace{0.2cm}\cellcolor{gray!25}64.10  &       &  \\ [2ex]

    \hline
    \hline
    \multirow{4}[8]{*}{\rotatebox{90}{\hspace{-0.3cm}\textbf{DI-NIDS}}} & \multirow{2}[4]{*}{\tiny{NFv2-CIC-2018}} & \vspace{0.2cm}{\tiny{NFv2-CIC-2018}} & \vspace{0.2cm}93.23 & \multirow{2}[4]{*}{\textbf{7.44}} & \multirow{4}[8]{*}{\textbf{6.42}} \\[2ex]
\cline{3-4}          &       & \vspace{0.2cm}\cellcolor{gray!25}{\tiny{NFv2-UNSW-NB15}} & \vspace{0.2cm}\cellcolor{gray!25}85.79 &       &  \\[2ex]
\cline{2-5}          & \multirow{2}[4]{*}{\tiny{NFv2-UNSW-NB15}} & \vspace{0.2cm}{\tiny{NFv2-UNSW-NB15}} & \vspace{0.2cm}98.68 & \multirow{2}[4]{*}{\textbf{5.39}} &  \\[2ex]
\cline{3-4}          &       & \vspace{0.2cm}\cellcolor{gray!25}{\tiny{NFv2-CIC-2018}} & \vspace{0.2cm}\cellcolor{gray!25}93.29 &       &  \\[2ex]
    \hline
    \end{tabular}%
  \label{tab:state_of_art}%
\end{table}%


\setlength{\textfloatsep}{15pt plus 2.0pt minus 4.0pt}
\subsection{\textbf{Comparison to Baseline Results}}
While there is no previous work evaluating the cross-domain performance of NIDSs using the same set of datasets as used in this study, there is a previous work~\cite{Pontes2021} running similar experiments using other datasets.
The authors of~\cite{Pontes2021} used two versions of the CIC-IDS dataset in their original format~\cite{cic}, CIC-2017 and CIC-2019, for their cross-domain evaluations.
Although we cannot compare the absolute performance numbers, we believe it is possible to compare the corresponding degree of performance degradation from the domain-specific to the cross-domain evaluation scenario.

Table~\ref{tab:state_of_art} shows the performances of DI-NIDS compared to the relevant state-of-the-art~\cite{Pontes2021}.
Here, each model is independently trained on two datasets, and evaluated on the training dataset (domain-specific evaluation), as well as the other, unseen dataset (cross-domain evaluation). The cross-domain results are shown as shaded cells in the table. 
The first column indicates the model, followed by the training dataset and the test dataset. The fourth column shows the corresponding F1-Scores.  

The fifth column presents the \textit{cross-domain degradation}, i.e. the difference between the F1-Score achieved in the domain-specific evaluation and the corresponding cross-domain evaluation. Finally, the sixth column shows the cross-domain degradation average across the two test and train dataset combinations.

%
%

While the absolute performance numbers of DI-NIDS and ~\cite{Pontes2021} in the table might not be comparable due the the use of different datasets, we argue that it is reasonable to compare the corresponding cross-domain degradation figures.
We can see that the cross-domain degradation values of DI-NIDS are 7.44\% and 5.39\% for the two 'directions' of evaluation, compared to 11.10\% and 27.50\% of~\cite{Pontes2021}. 

One of the benefits of DI-NIDS is that its performance degradation is relatively consistent across the two directions of cross-domain evaluation, with only a (absolute) difference of 2\% between the two values. 
In contrast, the the model proposed in ~\cite{Pontes2021} is sensitive to the direction of cross-domain evaluation, and exhibits a higher degree of variance, with difference of more than 16\% between the two directions of evaluation. 

Furthermore, and more importantly, the average cross-domain degradation of DI-NIDS is only 6.42\%, compared to 19.3\% to the relevant state-of-the-art ~\cite{Pontes2021}. This is a significant improvement in terms of cross-domain performance, and represents a step towards more domain invariant, and hence practical ML-based NIDSs.

\section{Conclusion}\label{conclusion}

This paper proposes a \textit{domain-invariant} network intrusion detection system (NIDS) framework to address the shortcomings of the existing NIDSs in regards to distribution shifts in data.
While domain adaptation methods to address the problem of distribution shift have been extensively studied in a range of machine learning application areas, it has not received significant attention in the context of ML-based NIDSs. 
This is despite the fact there are likely to be significant distribution shifts between training datasets and test data, in particular between (synthetic) training datasets and  real-world production networks, such as shown in~\cite{layeghy2021benchmarking}.
As we demonstrate in this paper, standard domain adaptation (DA) methods  based on a supervised learning approach do not work very well for NIDSs. One of the key reasons is that the existing DA approaches generally assume balanced datasets. Realistic datasets (and network traffic in general) in the context of NIDS are highly imbalanced, with attack or anomalous traffic representing only a small proportion of the overall network traffic. 

In order to address this gap, this paper proposes DI-NIDS, a domain invariant NIDS framework that takes into account the highly imbalanced nature of network traffic, while efficiently addressing the distribution shift between source and target domains.  
DI-NIDS achieves this by using a Domain-Adversarial Neural Network (DANN) to project the data into a domain-invariant feature space. 
The DANN is trained using data and labels from the source domain, and unlabelled data from the target domain.
DI-NIDS learns features that discriminate between classes in the source domain, but that do not discriminate between the source and target domains.
It leverages the feature extractor network of the trained DANN, and uses an OSVM model for the downstream task of traffic classification and anomaly detection.
Our experimental results show that, in addition to achieving excellent domain-specific classification performance, DI-NIDS significantly improves the cross-domain performance over the relevant state-of-the-art.  
We believe that improving the domain invariance, and robustness against feature distribution shifts, for ML-based NIDSs is an important step towards a more widespread deployment of such systems in practical real-world networks.

\section{Acknowledgement}
This research is made possible by an Advance Queensland Industry Research Fellowship, grant number RM2019002409. 


\bibliography{main}
\end{sloppypar}

\end{document}